\documentclass[letterpaper,11pt]{article}

\usepackage{amssymb}

\newcommand {\beq} {\begin{equation}}
\newcommand {\eeq} {\end{equation}}
\newcommand {\beqa} {\begin{eqnarray}}
\newcommand {\eeqa} {\end{eqnarray}}
\newcommand {\beqan} {\begin{eqnarray*}}
\newcommand {\eeqan} {\end{eqnarray*}}

\newcommand {\nn} {\nonumber}
\newcommand {\ph}[1]{\phantom{#1}}
\newcommand {\ie}{i.e.~}

\newcommand {\sss} {\scriptscriptstyle}

\newcommand{\al}{\ensuremath{\alpha}}
\newcommand{\be}{\ensuremath{\beta}}
\newcommand{\ga}{\ensuremath{\gamma}}
\newcommand{\Ga}{\ensuremath{\Gamma}}
\newcommand{\de}{\ensuremath{\delta}}

\newcommand{\eps}{\ensuremath{\epsilon}}

\newcommand{\ka}{\ensuremath{\kappa}}

\newcommand{\Si}{\ensuremath{\Sigma}}

\newcommand{\Om}{\ensuremath{\Omega}}
\newcommand{\Th}{\ensuremath{\Theta}}

\newcommand{\mathHb}[1]{{\mathop{\kern0pt#1}\limits^{\,\sss
      \prime\prime}\vphantom{#1}}}

\newcommand {\pa} {\partial}

\begin{document}

 \pagestyle{empty}
 \vskip-10pt
 \hfill {\tt hep-th/0402187}

\begin{center}

\vspace*{2cm}

% \rule{150mm}{2pt}
%\vskip 0.5truecm
\noindent
{\LARGE\textsf{\textbf{The $(2, 0)$ supersymmetric theory of
      tensor\\[5mm] multiplets and self-dual strings in six
      dimensions}}}

\vskip 1truecm
% \rule{150mm}{2pt}
\vskip 2truecm

{\large \textsf{\textbf{P\"ar Arvidsson\footnote{\tt par@fy.chalmers.se}, Erik
  Flink\footnote{\tt erik.flink@fy.chalmers.se} and M{\aa}ns
  Henningson\footnote{\tt mans@fy.chalmers.se}}}} \\ 
\vskip 1truecm
{\it Department of Theoretical  Physics\\ Chalmers University of
  Technology and G\"oteborg University\\ SE-412 96 G\"{o}teborg,
  Sweden}\\
\end{center}
\vskip 1cm
\noindent{\bf Abstract:}
We construct a unique $(2, 0)$ supersymmetric action in six
dimensions, describing a tensor multiplet interacting with a self-dual
string. It is a sum of four terms: A free kinetic term for the tensor
multiplet fields integrated over Minkowski space, a Nambu-Goto type
kinetic term for the string integrated over the string world-sheet, a
Wess-Zumino type electromagnetic coupling integrated over the
world-volume of a Dirac membrane attached to the string, and a direct
interaction of such Dirac membranes. In addition to supersymmetry, the
action is also invariant under a local symmetry, which allows us to
choose the Dirac membrane world-volume freely and eliminate half of
the fermionic degrees of freedom on the string world-sheet.

\newpage
\pagestyle{plain}

\section{Introduction} 
The six-dimensional $(2, 0)$ theories are one of the most remarkable
and surprising discoveries in string theory during the past decade
\cite{Witten95}. The existence of these theories can be inferred by
studying string theory or $M$-theory on a ten- or eleven-dimensional
space-time which contains a six-dimensional impurity. Under certain
conditions, degrees of freedom supported at the locus of this impurity
then decouple from the remaining bulk degrees of freedom and could be
described by a separate six-dimensional quantum theory. However, an
intrinsically six-dimensional definition of these theories is not yet
known, and finding such a description appears as a major goal of our
work. At first sight, it then appears natural to work at the origin of
the moduli space, but here a $(2, 0)$ theory is described by a
strongly coupled conformal field theory, which is hard to understand
with current methods. In previous papers
\cite{Arvidsson-Flink-Henningson}, we have instead followed an
approach based on working at a generic point in the moduli space,
where conformal invariance is spontaneously broken, and the theory can
be thought of as describing the dynamics of massless fields coupled to
tensile strings.

In this paper, we will construct an interacting theory of such
massless fields and tensile strings. We will limit ourselves to the
simplest $A_1$ version of $(2, 0)$ theory, where there is a single
species of fields and a single species of strings. Although we have
taken steps in this direction in our previous publications, we have
strived to make the present paper self-contained. In the next section,
we will review the free kinetic terms for the fields in
six-dimensional Minkowski space. The main difficulty here is that in
order to give a Lagrangian description, we must include additional
degrees of freedom that are not really part of $(2, 0)$ theory
\cite{Witten97}. This is consistent though, provided that the
interactions that we introduce later are constructed so that these
extra degrees of freedom remain decoupled. In section three, we will
discuss the corresponding on-shell superspace
\cite{Howe-Sierra-Townsend}. Although this is really only appropriate
for the free theory, it will still be an important tool for
constructing the interactions. In section four, we couple a string to
a prescribed background of space-time fields that fulfill the free
equations of motion. There are two interaction terms: One is a
Nambu-Goto type term for the string, with the tension given by a
space-time scalar field. The other is a Wess-Zumino type term that
encodes the electromagnetic coupling of the string to a space-time
tensor field. The Wess-Zumino term is constructed using a Dirac
membrane, the boundary of which is given by the string. In addition to
being supersymmetric, the sum of these two terms is invariant under a
local symmetry, which has two important consequences. First of all it
allows us to choose the Dirac membrane freely, so that only the actual
string is of physical significance
\cite{Deser-Gomberoff-Henneaux-Teitelboim}. Furthermore, this symmetry
decouples half of the fermionic variables on the string world-sheet,
reflecting the BPS-saturated property of the string
\cite{Bergshoeff-Sezgin-Townsend}. In the last section we will combine
the Minkowski space kinetic terms with the Nambu-Goto and Wess-Zumino
terms. To ensure that the complete action still is invariant under
supersymmetry and the local symmetry, we have to modify the
transformation laws of the space-time fields with additional terms. We
also have to add an additional term to the action describing direct
Dirac membrane interactions. The exact description of this term is
somewhat implicit, but we also show how its purely bosonic part can be
constructed completely explicitly.

The final action that we have constructed is unique, with no
adjustable parameters. This seems to indicate that the resulting
quantum theory, when properly defined, may have some finiteness
property. We hope to be able to investigate these questions in
forthcoming publications. 

\section{The tensor multiplet term}
We will begin by explaining our notation for $(2, 0)$ supersymmetry in
six-dimensional Minkowski space $M$. This symmetry of course contains
the Lorentz group $SO (5, 1)$. We denote chiral and anti chiral
Lorentz spinors by a subscript or superscript index $\al, \be, \ldots
= 1, 2, 3, 4$ respectively. All other representations may be obtained
by taking tensor products of these. We will have use for the
completely antisymmetric invariant tensors $\eps_{\al \be \ga \de}$
and $\eps^{\al \be \ga \de}$, defined so that $\eps_{\al \ga \de \eps}
\eps^{\be \ga \de \eps} = 6 \de_\al^\be$. Another bosonic subgroup of
$(2, 0)$ supersymmetry is an $SO (5)$ $R$-symmetry group. We denote
its spinor representation, from which all other representations can be
constructed, by an index $a, b, \ldots = 1, 2, 3, 4$. The
anti-symmetric invariant tensors are denoted $\Om^{a b}$ and $\Om_{a
  b}$ defined so that $\Om_{a b} \Om^{b c} = \de_a^c$.

The $(2, 0)$ supersymmetry algebra has a well-known representation on
a free field theory known as a tensor multiplet
\cite{Howe-Sierra-Townsend}. We denote the bosonic fields as
$b_\al^\be$ and $\phi^{a b}$, subject to the algebraic constraints
$b_\al^\al = 0$, $\phi^{a b} = - \phi^{b a}$ and $\phi^{a b} \Om_{a b}
= 0$. This means that $b_\al^\be$ is a two-form, and that the scalars
$\phi^{a b}$ transform in the vector representation of the
$R$-symmetry group. These fields are real. (This means that the
components $b_\al^\be$ and $\phi^{a b}$ are related to the complex
conjugates of other components by certain linear conditions.) The
fermionic fields are denoted as $\psi_\al^a$, \ie it is a chiral
spinor in the spinor representation of the $R$-symmetry group. This
field obeys a symplectic Majorana reality condition.

The field $b_\al^\be$ is subject to a gauge-invariance with a one-form
parameter $a_{\al \be} = - a_{\be \al}$, which acts as $b_\al^\be
\rightarrow b_\al^\be + \pa_{\al \ga} a^{\ga \be} + \frac{1}{4}
\de_\al^\be \pa_{\ga \de} a^{\ga \de}$. Here $\pa_{\al \be} = -
\pa_{\be \al}$ is the ordinary derivative acting on the space-time
coordinates $x^{\al \be} = - x^{\be \al}$. It is convenient to also
introduce $\pa^{\al \be} = \frac{1}{2} \eps^{\al \be \ga \de} \pa_{\ga
  \de}$. The gauge invariant information of $b_\al^\be$ is contained
in the self-dual and anti self-dual three-form field strengths $h_{\al
  \be}$ and $h^{\al \be}$ defined as 
\beqa
h_{\al \be} & = & \pa_{\al \ga} b_\be^\ga + \pa_{\be \ga} b_\al^\ga \cr
h^{\al \be} & = & \pa^{\al \ga} b_\ga^\be + \pa^{\be \ga} b_\ga^\al . 
\eeqa
They obey the Bianchi identity $\pa^{\al \ga} h_{\al \be} - \pa_{\al \be} h^{\al \ga} = 0$.

The dynamics of the fields $b_\al^\be$, $\phi^{a b}$ and $\psi_\al^a$ is described by the tensor multiplet action
\beq
S_{TM} = \int_M d^6 x \left( 2 h_{\al \be} h^{\al \be} - \Om_{a c} \Om_{b d} \pa_{\al \be} \phi^{a b} \pa^{\al \be} \phi^{c d}  - 4 i \Om_{a b} \psi^a_\al \pa^{\al \be} \psi^b_\be \right) .
\eeq
Under an arbitary variation of the fields, the variation of the action is
\beq
\de S_{TM} = \int_M d^6 x \left( 8 \de b_\al^\be \pa^{\al \ga} h_{\be \ga} + 2 \de \phi^{a b} \Om_{a c} \Om_{b d} \pa^{\al \be} \pa_{\al \be} \phi^{c d} - 8 i \de \psi^a_\al \Om_{a b} \pa^{\al \be} \psi^b_\be \right) , 
\eeq
from which the classical equations of motion can be read off:
\beqa \label{EOM}
\pa^{\al \ga} h_{\be \ga} & = & 0 \cr
\pa^{\al \be} \pa_{\al \be} \phi^{a b} & = & 0 \cr
\pa^{\al \be} \psi_\be^b & = & 0 .
\eeqa
Apart from its manifest Lorentz and $R$-symmetries, this action is also invariant under supersymmetries with constant infinitesimal fermionic parameters $\eta_a^\al$ acting as
\beqa \label{SUSY}
\de \phi^{a b} & = & -i \eta_c^\ga \left( \Om^{c a} \psi^b_\ga + \Om^{b c} \psi^a_\ga + \frac{1}{2} \Om^{a b} \psi^c_\ga \right) \cr
\de \psi^a_\al & = & \eta_c^\ga \left(\Om^{a c} h_{\al \ga} + 2 \pa_{\al \ga} \phi^{a c} \right) \cr
\de b_\al^\be & = & - i \eta^\be_a \psi_\al^a + \frac{i}{4} \de_\al^\be \eta^\ga_c \psi^c_\ga .
\eeqa
The transformation law for $b_\al^\be$ implies that the field strengths $h_{\al \be}$ and $h^{\al \be}$ transform as
\beqa
\de h_{\al \be} & = & i \eta_c^\ga \left( \pa_{\ga \al} \psi_\be^c + \pa_{\ga \be} \psi_\al^c \right) \cr
\de h^{\al \be} & = & - i \eta^\al_a \pa^{\be \ga} \psi_\ga^a - i \eta^\be_a \pa^{\al \ga} \psi_\ga^a .
\eeqa

We note that the anti self-dual part $h^{\al \be}$ of the field
strength of $b_\al^\be$ does not appear in the right hand sides of
these supersymmetry transformation laws. Furthermore, the
supersymmetry variation $\de h^{\al \be}$ vanishes when the equations
of motion of the field $\psi^a_\al$ is imposed. These two properties
imply that $h^{\al \be}$ is not really part of the tensor multiplet,
which thus consists solely of the fields $\phi^{a b}$, $\psi_\al^a$,
and $h_{\al \be}$. It is well-known that a covariant action describing
only a tensor multiplet does not exist. The above action describes a
free tensor multiplet and a free anti self-dual field strength $h^{\al
  \be}$. Later in this paper, we will construct interaction terms that
couple the tensor multiplet to a string. We will however make sure
that $h^{\al \be}$ remains decoupled from all other degrees of
freedom.

The fields $\phi^{a b}$ and $\psi^a_\al$ take their values in linear
spaces, so the numerical coefficients in their kinetic terms in
$S_{TM}$ are purely conventional, and just reflects the normalization
of these fields. But the field $b_\al^\be$ is really a connection on a
1-gerbe, so its normalization cannot be changed. The constant in front
of its kinetic term in $S_{TM}$ is therefore significant. A priori, it
is a free parameter of the theory, but by considering the theory on a
space-time of non-trivial topology, one finds that the decoupling of
the anti self-dual part $h^{\al \be}$ of the field strength is only
consistent for a specific value of this constant. (In this paper, we
will make no attempt to define our conventions precisely enough to
determine this value.) So the action $S_{TM}$ is in fact unique, with
no free parameters.

\section{The on-shell superspace of a free tensor multiplet}
Although the aim of this paper is to construct an interacting theory, for which no superspace formulation is known, it is still useful to introduce the on-shell superspace of the free tensor multiplet described in the previous section \cite{Howe-Sierra-Townsend}.  

Superspace is parametrized by the bosonic coordinates $x^{\al \be}$ of six-dimensional Minkowski space together with a set of fermionic coordinates $\theta^\al_a$, which are symplectic Majorana. Supersymmetry transformations are generated by the supercharges
\beq
Q_\al^a = \pa_\al^a - i \Om^{a b} \theta^\be_b \pa_{\al \be} ,
\eeq
where $\pa_\al^a = \frac{\pa}{\pa \theta_a^\al}$. These anti-commute with the covariant derivatives
\beq
D_\al^a = \pa_\al^a + i \Om^{a b} \theta^\be_b \pa_{\al \be} .
\eeq
We will also need differential forms on superspace. As a basis of one-forms, we will use $e^{\al \be} = -e^{\be \al}$ and $d \theta^\al_a$, where 
\beq
e^{\al \be} = d x^{\al \be} + \frac{i}{2} \Om^{a b} (\theta^\al_a d \theta^\be_b - \theta^\be_a d \theta^\al_b) .
\eeq
These are dual to the tangent vectors $\pa_{\al \be}$ and $D_\al^a$ respectively, so that the exterior derivative $d$ takes the form
\beq
d = e^{\al \be} \pa_{\al \be} + d \theta^\al_a D_\al^a .
\eeq
It follows that $d e^{\al \be} = i \Om^{a b} d \theta ^\al_a \wedge d \theta^\be_b$, whereas $d \theta$ of course is closed. 

We now introduce a superfield $\Phi^{a b} (x, \theta)$, subject to algebraic constraints analogous to those of $\phi^{a b} (x)$, \ie $\Phi^{a b} = - \Phi^{b a}$ and $\Om_{a b} \Phi^{a b} = 0$. It should also fulfill the differential constraint
\beq \label{constraint}
D^a_\al \Phi^{b c} + \frac{1}{5} \Om_{d e} D^d_\al \left(2 \Om^{a b} \Phi^{e c} - 2 \Om^{a c} \Phi^{e b} + \Om^{b c} \Phi^{e a} \right) = 0 .
\eeq
To analyse the implications of these constraints, it is convenient to define the superfields $\Psi^a_\al$ and $H_{\al \be}$ as
\beqa \label{PsiH}
\Psi^a_\al & = & -\frac{2 i}{5} \Om_{b c} D^b_\al \Phi^{c a} \cr
H_{\al \be} & = & \frac{1}{4} \Om_{a b} D^a_\al \Psi^b_\be .
\eeqa
It immediately follows that $H_{\al \be}$ obeys the algebraic constraint $H_{\al \be} = H_{\be \al}$. Furthermore, the differential constraint on $\Phi^{a b}$ implies that all fermionic derivatives of $\Phi^{a b}$ may be expressed in terms of $\Phi^{a b}$, $\Psi^a_\al$, $H_{\al \be}$ and bosonic derivatives thereof. Notably
\beqa
D^a_\al \Phi^{b c} & = & - i \left( \Om^{a b} \Psi^c_\al + \Om^{c a} \Psi^b_\al + \frac{1}{2} \Om^{b c} \Psi^a_\al \right) \cr
D^a_\al \Psi^b_\be & = & 2 \pa_{\al \be} \Phi^{a b} - \Om^{a b} H_{\al \be} \cr
D^a_\al H_{\be \ga} & = & i (\pa_{\al \be} \Psi^a_\ga + \pa_{\al \ga} \Psi^a_\be) .
\eeqa
Finally, the differential constraint also implies the equations of motion
\beqa
\pa^{\al \ga} H_{\be \ga} & = & 0 \cr
\pa^{\al \be} \pa_{\al \be} \Phi^{a b} & = & 0 \cr
\pa^{\al \be} \Psi_\be^b & = & 0 .
\eeqa
We will also need a certain three-form $F$ on superspace
\cite{Grojean-Mourad,Howe} defined as
\beq \label{F}
F = -\frac{1}{12} e^{\al \be} \wedge e^{\ga \de} \wedge e^{\eps \ka} \eps_{\be \ga \de \eps} H_{\al \ka} - \frac{i}{4} e^{\al \be} \wedge e^{\ga \de} \wedge d \theta^\eps_a \eps_{\be \ga \de \eps} \Psi^a_\al + \frac{i}{4} e^{\al \be} \wedge d \theta^\ga_a \wedge d \theta^\de_b \eps_{\al \be \ga \de} \Phi^{a b} 
\eeq
Acting with the exterior derivative $d$ on $F$, and using the above properties of the superfields $\Phi^{a b}$, $\Psi^a_\al$, and $H_{\al \be}$, one finds that $F$ is closed, \ie $d F = 0$. 

The general solution to the above constraints on the superfield
$\Phi^{a b} (x, \theta)$ may be written in terms of component fields
$\phi^{a b} (x)$, $\psi^a_\al (x)$, and $h_{\al \be} (x)$, that are
the lowest components of the superfields $\Phi^{a b} (x, \theta),
\Psi^a_\al (x, \theta)$, and $H_{\al \be} (x, \theta)$
respectively. It follows that these component fields have to obey the
algebraic constraints and the free classical equations of motion of
the tensor multiplet fields discussed in the previous
section. Including terms up to bilinear order in $\theta$, one finds
that
\beqa \label{Phi}
\Phi^{a b} & = & \phi^{a b} - i \theta^\al_c \left(\Om^{c a}
\psi^b_\al + \Om^{b c} \psi^a_\al + \frac{1}{2} \Om^{a b} \psi^c_\al
\right) \cr
& & {} + i \theta^\al_c \theta^\be_d \left(h_{\al \be} (\Om^{d a}
\Om^{b c} + \frac{1}{4} \Om^{d c} \Om^{a b}) - \Om^{d a} \pa_{\al \be}
\phi^{b c} - \Om^{d b} \pa_{\al \be} \phi^{c a} \right) + {\cal O}
(\theta^3). \qquad
\eeqa
The coefficients of the neglected terms are given by derivatives of
the components fields $\phi^{a b}$, $\psi^a_\al$, and $h_{\al \be}$,
and the expansion ends with a term of order $\theta^{16}$. Acting with
a supersymmetry transformation, generated by the supercharges, one
finds that the transformation laws of the component fields agree with
those presented in the previous section, thus completing their
identification with the tensor multiplet fields.

\section{The Nambu-Goto and Wess-Zumino terms}
In this section, we will couple a string to a fixed tensor multiplet
background, that fulfills the free equations of motion described in
section two. This is an intermediate step, before constructing the
complete dynamical theory of tensor multiplets and strings in the next
section.

On physical grounds, we expect two types of couplings to appear: The
first is a Nambu-Goto term given by an integral over the string
world-sheet $\Sigma$. The integrand is the volume form on $\Sigma$,
induced by its embedding into Minkowski space, times the $R$-symmetry
invariant norm of the scalar fields $\phi^{a b}$. This
term thus reflects the fact that the tension of the string is
determined by the moduli. The second term is a Wess-Zumino term given
by an integral over a Dirac membrane world-volume $D$, the boundary
$\pa D$ of which equals the string world-sheet $\Sigma$. The integrand
is the pullback of the self-dual field strength $h_{\al \be}$ to
$D$. This term thus encodes the electromagnetic coupling of the
self-dual string to the field $b_\al^\be$. Our problem is now to
construct the supersymmetric versions of these couplings.

The embedding of the Dirac membrane world-volume $D$ in superspace is described by world-volume fields $X^{\al \be}$ and $\Th^\al_a$. They transform non-linearly under supersymmetry according to
\beqa
\de X^{\al \be} & = & \frac{i}{2} \Om^{a b} \left( \eta^\al_a \Th^\be_b - \eta^\be_a \Th^\al_b \right) \cr
\de \Th^\al_a & = & - \eta^\al_a .
\eeqa
It follows that the pullbacks to $D$ of the one-forms $e^{\al \be}$ and $d \theta^\al_a$ introduced in the previous section, \ie $E^{\al \be}$ given by
\beq
E^{\al \be} = d X^{\al \be} + \frac{i}{2} \Om^{a b} (\Th^\al_a d \Th^\be_b - \Th^\be_a d \Th^\al_b) 
\eeq
and $d \Th^\al_a$, are invariant under supersymmetry. If we parametrize the string world-sheet $\Sigma$ with coordinates $\sigma^i$, $i = 1, 2$ and expand the one-forms $E^{\al \be}$ in the basis $d \sigma^i$ as $E^{\al \be} = d \sigma^i E_i^{\al \be}$, we can construct a supersymmetric induced metric $G_{i j}$ on $\Sigma$ as
\beq
G_{i j} = \frac{1}{4} \eps_{\al \be \ga \de} E_i^{\al \be} E_j^{\ga \de} .
\eeq
We denote the determinant of this metric as $G$.

Supersymmetry transformations on the superfield $\Phi^{a b}$ are generated by the supercharges $Q^a_\al$. It follows that the pullback ${}^* \Phi^{a b} = \Phi^{a b} (X, \Th)$ of $\Phi^{a b}$ to $D$ is invariant under supersymmetry, since the contributions to its variation from the two terms in $Q^a_\al$ are precisely cancelled by the contributions due to the variations of its arguments $X$ and $\Theta$. It is then straightforward to construct a supersymmetric Nambu-Goto term $S_{NG}$ as
\beq
S_{NG} = - \int_\Sigma d^2 \sigma \sqrt{{}^* (\Phi \cdot \Phi)} \sqrt{- G} ,
\eeq
where $\Phi \cdot \Phi$ is defined as
\beq
\Phi \cdot \Phi = \frac{1}{4} \Om_{a c} \Om_{b d} \Phi^{a b} \Phi^{c d} .
\eeq
Similarly, the pullbacks of the superfields $\Psi^a_\al$ and $H_{\al \be}$ to $D$ are invariant under supersymmetry, and so are the differentials $E^{\al \be}$ and $d \Th^\al_a$. We may thus construct a supersymmetric Wess-Zumino term $S_{WZ}$ as
\beq
S_{WZ} = \int_D {}^* F ,
\eeq
where ${}^* F$ denotes the pullback to $D$ of the three-form $F$ introduced in the previous section.

The interaction terms must be invariant under a change of the Dirac membrane world-volume $D$, as long as its boundary is given by the string world-sheet $\Sigma$. An infinitesimal change of $D$ is parametrized by a superspace tangent vector field $\ka$ on $D$. Expanding this as $\ka = \ka^{\al \be} \pa_{\al \be} + \ka^\al_a D^a_\al$, we get the transformation laws
\beqa
\de X^{\al \be} & = & \ka^{\al \be} + \frac{i}{2} \Om^{a b} \left( \ka^\al_a \Th^\be_b - \ka^\be_a \Th^\al_b \right) \cr
\de \Th^\al_a & = &  \ka^\al_a ,
\eeqa
for the world-volume fields $X^{\al \be}$ and $\Th^\al_a$ that describe the embedding of $D$ in superspace. The superfield $\Phi^{a b}$ is of course invariant under this transformation. To describe a change which leaves the boundary $\Sigma$ fixed, $\ka$ should vanish there. We will, however, be slightly more general and only require that the bosonic components $\ka^{\al \be}$ vanish on $\Sigma$, whereas the fermionic components $\ka^\al_a$ on $\Sigma$ are subject to the constraint 
\beq
\Ga^\al{}_\be \kappa^\be_a = \ga_a{}^b \kappa^\al_b .
\eeq
Here $\Ga^\al{}_\be$ and $\ga_a{}^b$ are defined as
\beqa
\Ga^\al{}_\be & = & \frac{1}{2} \frac{1}{\sqrt{-G}} \eps^{i j} E_i^{\al \ga} E_j^{\de \eps} \eps_{\be \ga \de \eps} \cr
\ga_a{}^b & = & \frac{1}{\sqrt{{}^* (\Phi \cdot \Phi})} \Om_{a c} {}^* \Phi^{c b} ,
\eeqa
and fulfill the identities $\Ga^\al{}_\be \Ga^\be{}_\ga = \de^\al_\ga$, $\ga_a{}^b \ga_b{}^c = \de_a^c$, $\Ga^\al{}_\al = 0$, and $\ga_a{}^a = 0$. The constraint on $\ka^\al_a$ on $\Sigma$ thus means that it has eight linearly independent components, so that eight of the sixteen components of $\Th^\al_a$ may be eliminated by a $\ka$-symmetry transformation. The remaining eight components of $\Th^\al_a$ correspond to four fermionic degrees of freedom. This equals the number of bosonic degrees of freedom, after eliminating two components of $X^{\al \be}$ by invariance under reparametrizations of $\Sigma$. We remark that in many cases that have appeared in the literature, the integrand of the Wess-Zumino term is a total derivative, so by Stokes' theorem this term can be rewritten as an integral over the boundary $\Sigma$. The relevant part of the $\ka$-symmetry is then of course parametrized by the fermionic components $\ka^\al_a$ subject to the above constraint.

It remains to show that the sum of the interaction terms $S_{WZ}$ and $S_{NG}$ is indeed invariant under the transformation parametrized by the vector field $\ka$ on $D$ described above. Let $\omega$ be an arbitary differential form on superspace, invariant under $\ka$-symmetry. Its pullback ${}^* \omega$ to $D$ then transforms as 
\beq \label{kappa_pullback}
\de \, {}^* \omega ={}^* (\iota_\ka d \omega) +  d \, {}^* (\iota_\ka \omega) ,
\eeq
where $\iota_\ka$ denotes contraction with the tangent vector $\ka$. When this formula is applied to $\omega = F$, the first term vanishes since $F$ is closed. Stokes' theorem then gives the variation of the Wess-Zumino term as
\beq
\de S_{WZ} = \int_\Sigma {}^*(\iota_\ka F) = \int_\Sigma
\ka^\al_a \frac{i}{4} \epsilon_{\al \be \ga \de}\!\! {\ph{\Big(}}^*\!\! \left(
e^{\be \ga} \wedge e^{\de \eps} \Psi^a_\eps - 2 e^{\be \ga} \wedge d
\theta^\de_b \Phi^{a b} \right),
\eeq
where we have used the vanishing of the bosonic components of $\ka$ on $\Sigma$ in the last equality. 

The variation of the Nambu-Goto term is given by
\beq
\de S_{NG} = - \int_\Sigma d^2 \sigma \left\{\frac{\sqrt{-G}}{\sqrt{{}^* (\Phi \cdot \Phi)}} ({}^* \Phi \cdot \de {}^* \Phi) + \sqrt{{}^* (\Phi \cdot \Phi)} \frac{\de (-G)}{2 \sqrt{-G}} \right\} .
\eeq
Here the transformation of the pullback of $\Phi^{a b}$ is given by
$\de \, {}^* \Phi^{a b} = {}^* (\iota_\ka d \Phi^{a b}) = i \ka^\al_c
\; {}^* (\Om^{a c} \Psi_\al^b - \Om^{b c} \Psi_\al^a - \frac{1}{2}
\Om^{a b} \Psi_\al^c)$, where we have expressed the fermionic
derivatives of $\Phi^{a b}$ in terms of $\Psi^a_\al$ and used the
vanishing of the bosonic components of $\ka$. Finally, we should use
the constraints on $\ka^a_\al$ that are valid on $\Sigma$. One then
easily finds that the first term in $\de S_{WZ}$ cancels against the
first term in $\de S_{NG}$. A trickier computation shows that the
second term in $\de S_{WZ}$ cancels against the second term in
$S_{NG}$. The sum $S_{NG} + S_{WZ}$ is thus invariant under
$\ka$-symmetry.

We have seen that the relative coefficient between the Nambu-Goto and
Wess-Zumino terms is determined by the requirement of
$\kappa$-symmetry. But since the Wess-Zumino term represents an
electromagnetic coupling of the string to the field $b_\al^\be$, its
coefficient must obey Dirac quantization, \ie it is given by an
integer in appropriate units. We expect our strings to have the
minimal non-trivial charge, which thus fixes this value completely. So
there are no free parameters in the linear interaction terms.

\section{The Dirac-Dirac term}
In this section, we will describe how to construct the complete
dynamical theory of an interacting tensor multiplet and a string. The
main problem is as follows: We still define $\Phi^{a b}$,
$\Psi^a_\al$, $H_{\al \be}$, and $F$ in terms of the tensor multiplet
fields $\phi^{a b}$, $\psi^a_\al$, and $h_{\al \be}$ precisely as
before using equations (\ref{Phi}), (\ref{PsiH}), and (\ref{F}). But
crucial properties of these quantities are only valid when the tensor
multiplet fields obey their free equations of motion (\ref{EOM}), and
thus cannot be used in the interacting theory. Indeed, for generic
configurations of the tensor multiplet fields, these quantitites do
not transform as superfields, so the interaction terms $S_{NG}$ and
$S_{WZ}$ are not invariant under supersymmetry. Furthermore, the
fermionic derivatives of $\Phi^{a b}$ are no longer related by the
differential constraint (\ref{constraint}), \ie they cannot all be
expressed in terms of $\Psi^a_\al$, and this invalidates the
conclusion that $F$ is closed. Thus the sum $S_{NG} + S_{WZ}$ is not
invariant under $\ka$-symmetry. In this section we will show how these
problems can be remedied by supplementing the tensor multiplet action
$S_{TM}$ and the interaction terms $S_{NG}$ and $S_{WZ}$ with a direct
Dirac membrane interaction $S_{DD}$. The transformation laws of the
tensor multiplet fields under supersymmetry and $\ka$-symmetry also
have to be modified, by including terms supported on the Dirac
membrane world volume $D$ and its boundary $\Sigma$.

\subsection{Off-shell $\kappa$-symmetry}
The interaction $S_{NG} + S_{WZ}$ is invariant under a $\kappa$-symmetry transformation as described in the previous section, provided that the tensor multiplet fields $\phi^{a b}$, $\psi^a_\al$, and $h_{\al \be}$ fulfill their free equations of motion. Its variation when no such on-shell constraints are imposed must therefore be of the form
\beq
\de (S_{NG} + S_{WZ}) = \int_M d^6 x \left( -8 \check{b}_\al^\be \pa^{\al \ga} h_{\be \ga} - 2 \check{\phi}^{a b} \Om_{a c} \Om_{b d} \pa^{\al \be} \pa_{\al \be} \phi^{c d} + 8 i \check{\psi}^a_\al \Om_{a b} \pa^{\al \be} \psi^b_\be \right) , 
\eeq
for some coefficient functions $\check{b}_\al^\be$, $\check{\phi}^{a b}$, and $\check{\psi}^a_\al$ that are linear in the parameters $\ka$ of the transformation. This variation can thus be cancelled by a variation $\de_1 S_{TM}$ of the tensor multiplet action, \ie
\beq
\de (S_{NG} + S_{WZ}) + \de_1 S_{TM} = 0 ,
\eeq
if we let the tensor multiplet fields transform as
\beqa
\de_1 b_\al^\be & = & \check{b}_\al^\be \cr
\de_1 \phi^{a b} & = & \check{\phi}^{a b} \cr
\de_1 \psi^a_\al & = & \check{\psi}^a_\al .
\eeqa
But the latter transformations lead to an additional variation $\de_1 (S_{NG} + S_{WZ})$ of the interaction. We hope to cancel this by adding an extra Dirac-Dirac term $S_{DD}$ to the action, constructed so that
\beq
\de_1 (S_{NG} + S_{WZ}) + \de S_{DD} = 0 .
\eeq
If furthermore $S_{DD}$ is independent of the tensor multiplet fields, and only depends on the world volume fields $X^{\al \be}$ and $\Th^\al_a$, the transformation laws of which have not been altered, then $\de_1 S_{DD} = 0$, and the complete action
\beq
S = S_{TM} + S_{NG} + S_{WZ} + S_{DD}
\eeq
is exactly invariant under the modified $\kappa$-symmetry.

To see that such a term $S_{DD}$ indeed exists, we will have to retrace the above steps a bit more carefully. Repeating the analysis of the previous section but retaining also terms proportional to the free equations of motion of the tensor multiplet fields, we find that
\beq
\de (S_{NG} + S_{WZ}) = \int_D {}^* (\iota_\ka d F) + \ldots .
\eeq
In this and subsequent formulas, $\ldots$ denotes terms that are supported on the string world-sheet $\Sigma$. The four-form $d F$ vanishes when the tensor multiplet fields fulfill their free equations of motion, so it can be written as 
\beqa
d F & = & \int_M d^6 x^\prime \Big( -8 \tilde{b}_\al^\be (x^\prime)
\pa^{\prime \al \ga} h_{\ga \be} (x^\prime) - 2 \tilde{\phi}^{a b}
(x^\prime) \pa^{\prime \al \be} \pa^\prime_{\al \be} \phi^{c d}
(x^\prime) \Om_{a c} \Om_{b d} + {} \nn \\
& & {} + 8 i \tilde{\psi}^a_\al (x^\prime)
\pa^{\prime \al \be} \psi^b_\beta (x^\prime) \Om_{a b} \Big). 
\eeqa
Here, $\tilde{b}_\al^\be (x^\prime)$, $\tilde{\psi}^a_\al (x^\prime)$, and $\tilde{\phi}^{a b} (x^\prime)$ are some four-forms on superspace with coordinates $x^{\al \be}$ and $\theta^\al_a$, but as indicated they also depend on the variables $x^{\prime \al \be}$. We note that they are closed (since $d F$ is closed), and independent of the tensor multiplet fields (since $d F$ is linear in these). The transformation laws of the tensor multiplet fields thus become
\beqa
\de_1 \phi^{a b} (x^\prime) & = & \int_D \!\! {\ph{\Big(}}^*\!\! \left(\iota_\ka \tilde{\phi}^{a b} \right) + \ldots \cr
\de_1 \psi^a_\al (x^\prime) & = & \int_D \!\! {\ph{\Big(}}^*\!\! \left(\iota_\ka \tilde{\psi}^a_\al \right) + \ldots \cr
\de_1 h_{\al \be} (x^\prime) & = & \int_D \!\! {\ph{\Big(}}^*\!\! \left(\iota_\ka \tilde{h}_{\al \be} \right) + \ldots ,
\eeqa
where we have defined the closed four-form $\tilde{h}_{\al \be} (x^\prime)$ as
\beq
\tilde{h}_{\al \be} (x^\prime) = \pa_{\al \ga}^\prime \tilde{b}^\ga_\be + \pa_{\be \ga}^\prime \tilde{b}^\ga_\al .
\eeq
Since the cohomology of superspace is trivial, the closed forms $\tilde{\phi}^{a b} (x^\prime)$, $\tilde{\psi}^a_\al (x^\prime)$, and $\tilde{h}_{\al \be} (x^\prime)$ are in fact exact and can thus be written as 
\beqa
\tilde{\phi}^{a b} (x^\prime) & = & d \hat{\phi}^{a b} (x^\prime) \cr
\tilde{\psi}^a_\al (x^\prime) & = & d \hat{\psi}^a_\al (x^\prime) \cr
\tilde{h}_{\al \be} (x^\prime) & = & d \hat{h}_{\al \be} (x^\prime)
\eeqa
with some three-forms $\hat{\phi}^{a b} (x^\prime)$, $\hat{\psi}^a_\al
(x^\prime)$, and $\hat{h}_{\al \be} (x^\prime)$ that are functionals
of the world-volume fields $X^{\al \be}$ and $\Th^\al_a$. The exterior
derivative $d$ on the right hand side of course acts on the superspace
coordinates $x^{\al \be}$ and $\theta^\al_a$. (The closed form
$\tilde{b}_\al^\be (x^\prime)$ is also exact, but we cannot express it
as the exterior derivative $d$ acting on some local functional of
$X^{\al \be}$ and $\Th^\al_a$.) We can now use the $\ka$-symmetry
transformation law (\ref{kappa_pullback}) for the pullback of a
differential form applied to the three-forms $\hat{\phi}^{a b}
(x^\prime)$, $\hat{\psi}^a_\al (x^\prime)$, and $\hat{h}_{\al \be}
(x^\prime)$. We thus find that the $\ka$-symmetry transformations of
the tensor multiplet fields are given by
\beqa
\de_1 \phi^{a b} (x^\prime) & = & \de \int_D {}^* \hat{\phi}^{a b} (x^\prime) + \ldots \cr
\de_1 \psi^a_\al (x^\prime) & = & \de \int_D {}^* \hat{\psi}^a_\al (x^\prime) + \ldots \cr
\de_1 h_{\al \be} (x^\prime) & = & \de \int_D {}^* \hat{h}_{\al \be} (x^\prime) + \ldots ,
\eeqa
where $\de$ on the right hand sides denotes the $\kappa$-symmetry variation acting on the world-volume fields $X^{\al \be}$ and $\Theta^\al_a$. Again, the $\ldots$ denote terms that are supported on the boundary $\Sigma$. For $\Phi^{a b} (x^\prime, \theta^\prime)$, we thus get the transformation law
\beq
\de_1 \Phi^{a b} (x^\prime, \theta^\prime) = \de \int_D {}^* \hat{\Phi}^{a b} (x^\prime, \theta^\prime) + \ldots , 
\eeq
where the three-form $\hat{\Phi}^{a b} (x^\prime, \theta^\prime)$ is constructed out of $\hat{\phi}^{a b} (x^\prime)$, $\hat{\psi}^a_\al (x^\prime)$ and $\hat{h}_{\al \be} (x^\prime)$ in the same way as $\Phi^{a b}$ is constructed out of $\phi^{a b}$, $\psi^a_\al$, and $h_{\al \be}$. The quantities $\Psi^{a b} (x^\prime, \theta^\prime)$ and $H_{\al \be} (x^\prime, \theta^\prime)$ transform analogously. Finally, we consider the three-form $F^\prime$ in the variables $x^{\prime \al \be}$ and $\theta^{\prime \al}_a$ obtained as the pullback of $F$ by the map $x^{\prime \al \be} \mapsto x^{\al \be}$, $\theta^{\prime \al}_a \mapsto \theta^\al_a$. It transforms as
\beq
\de_1 F^\prime = \de \int_D {}^* \hat{F}^\prime  + \ldots .
\eeq
Here, $\hat{F}^\prime$ is a three-form in $x^{\prime \al \be}$ and $\theta^{\prime \al}_a$ constructed out of $\hat{\Phi}^{a b}$, $\hat{\Psi}^a_\al$ and $\hat{H}_{\al \be}$ in the same way as $F$ is constructed out of $\Phi^{a b}$, $\Psi^a_\al$ and $H_{\al \be}$. It is thus also a three-form in the variables $x^{\al \be}$ and $\theta^\al_a$, on which the pullback map ${}^*$ to $D$ acts. 

We should now really compute the $\de_1$ variation of $S_{NG} +
S_{WZ}$. However, $S_{WZ}$ and $S_{NG}$ are given by integrals over
$D$ and its boundary $\Sigma$ respectively, and the $\de_1$ variations
of the tensor multiplet fields is a sum of terms supported on $D$ and
terms supported on $\Sigma$. We thus get a divergent expression. To
regularize this, we introduce another open three-manifold $D^\prime$
with boundary $\Sigma^\prime$ by slightly perturbing $D$. We can now
compute the $\de_1$ variation of the regularized interaction terms
$S_{NG}^\prime + S_{WZ}^\prime$ obtained by replacing $D$ and $\Sigma$
with $D^\prime$ and $\Sigma^\prime$ in $S_{NG}$ and $S_{WZ}$. For a
generic perturbation, the intersection of $D^\prime$ with $\Sigma$ and
the intersection of $\Sigma^\prime$ with $D$ are empty, while $D$ and
$D^\prime$ intersect in isolated points. This is analogous to the
Dirac veto familiar from the theory of magnetic monopoles in four
space-time dimensions, where the Dirac string should be chosen so that
it does not go through any particle, and furthermore the effects on a
particle of electromagnetic fields caused by that particle itself are
not considered. It is also analogous to the definition of the
self-intersection number of a submanifold.

These properties of $D^\prime$ and $\Sigma^\prime$ mean that $\de_1 S_{NG}^\prime$ vanishes, whereas terms in $\de_1 F^\prime$ that are supported on $\Sigma$ can be neglected in the computation of $\de_1 S_{WZ}^\prime$. Using the results of the previous paragraph, we thus find that 
\beq
\de_1 (S_{NG}^\prime + S_{WZ}^\prime) + \de S_{D^\prime D} = 0 ,
\eeq
if we define the regularization $S_{D^\prime D}$ of the new interaction term $S_{DD}$ as
\beq
S_{D^\prime D} = -\frac{1}{2} \int_{D^\prime} \int_D {}^* \, {}^{* \prime} \hat{F}^\prime .
\eeq
Here, ${}^{* \prime}$ denotes the pullback to $D^\prime$ acting on the variables $x^{\prime \al \be}$ and $\theta^{\prime \al}_a$. The factor $\frac{1}{2}$ arises because the variation $\de$ acts not only on the fields $X^{\al \be}$ and $\Th^\al_a$ on $D$, but also on the fields $X^{\prime \al \be}$ and $\Th^{\prime \al}_a$ on $D^\prime$, which transform analogously. We see that $S_{D^\prime D}$ represents a direct Dirac membrane interaction. As required, $S_{DD}$ is independent of the tensor multiplet fields. This is ultimately a consequence of the fact that the Wess-Zumino term is not only of first order in the tensor multiplet fields, like the Nambu-Goto term, but furthermore linear in them. This completes the proof of $\kappa$-symmetry of the complete action $S$.

\subsection{The bosonic theory}
There is no simple closed expression for the form $\hat{F}^\prime$ that appears in $S_{DD}$, but it can be constructed order by order as a polynomial in the fermionic variables $\theta^\al_a$ and $\theta^{\prime \al}_a$, just like for example the superfield $\Phi^{a b}$ is expanded in $\theta^\al_a$ in (\ref{Phi}). It is instructive, though, to show the result of this procedure to lowest order in fermions, \ie to determine the bosonic part $S^{\rm bos}$ of the complete action $S$.

For this purpose, it is convenient to employ an index-free notation: The bosonic world-volume fields are thus a map $X$ from $D$ to space-time $M$. The bosonic space-time fields are a two-form $b$ and an $R$-symmetry vector $\phi$ of scalars. The field strength $h = d b$ can be decomposed as $h = h_+ + h_-$, where $h_+ = \frac{1}{2} (h + * h)$ and $h_- = \frac{1}{2} (h - * h)$ are the self-dual and anti self-dual parts respectively. (The $*$ is the Hodge duality operator.) The bosonic parts of $S_{TM}$, $S_{NG}$, and $S_{WZ}$ are 
\beqa
S_{TM}^{\rm bos} & = & \frac{1}{2} \int_M h \wedge * h + \frac{1}{2} \int_M d \phi \cdot \wedge * d \phi \cr
S_{NG}^{\rm bos} & = & \int_\Sigma \sqrt{\phi \cdot \phi} \sqrt{-g} \cr
S_{WZ}^{\rm bos} & = & \int_D 2 h_+ .
\eeqa
(In this subsection, to avoid confusion with the Hodge duality operator $*$, we suppress the ${}^*$ that indicates the pullback of a differential form to $D$ or $\Sigma$.) Here $g$ denotes the determinant of the metric on $\Sigma$ induced by its embedding in six-dimensional Minkowski space. To proceed further, we need to introduce the Poincar\'e dual three-form $\de_D$ of the three-manifold $D$ embedded in six-dimensional Minkowski space, defined by the property that 
\beq
\int_M \de_D \wedge s = \int_D s
\eeq
for an arbitrary three-form $s$ on Minkowski space. It can be written as
\beq
\de_D = \left<d x \wedge d x \wedge d x  \right. \int_D \left. d X \wedge d X \wedge d X \right> \de^{(6)} (x - X) ,
\eeq
where $\left< \ldots \ldots \right>$ means that the three space-time differentials $d x$ and the three pullback differentials $d X$ are contracted with the six-dimensional Levi-Civita tensor. We can now write the bosonic part of $S_{DD}$ as
\beq
S_{DD}^{\rm bos} = \frac{1}{2} \int_{D} * \de_D .
\eeq
As described in the previous subsection, this expression is divergent since $* \de D$ is supported on $D$, and has to be regularized by replacing one copy of $D$ with a perturbation $D^\prime$. The complete bosonic action $S^{\rm bos} = S_{TM}^{\rm bos} + S_{NG}^{\rm bos} + S_{WZ}^{\rm bos} + S_{DD}^{\rm bos}$ can be rewritten as
\beq
S^{\rm bos} = \frac{1}{2} \int_M h^{\rm tot} \wedge * h^{\rm tot} + \int_M d \phi \cdot \wedge * d \phi + \int_\Sigma \sqrt{\phi \cdot \phi} \sqrt{-g} + \int_\Sigma b ,
\eeq
where the total field strength $h^{\rm tot}$ is defined as
\beq
h^{\rm tot} = h + \de_D .
\eeq
The last term in the action is obtained by applying Stokes' theorem to the term $\int_D h$ and represents the standard electric coupling of the string to the field $b$. The first term can then be understood as a way of incorporating the magnetic coupling. Indeed, $h^{\rm tot}$ fulfills a modified Bianchi identity $d h^{\rm tot} = \de_\Sigma$, where $\de_\Sigma = d \de_D$ is the Poincar\'e dual of the string world-sheet $\Sigma$. This is appropriate in the presence of a magnetically charged string. 

An infinitesimal bosonic $\kappa$-symmetry transformation is parametrized by a vector $\kappa$ on $D$, subject to the condition that it vanishes on $\Sigma$. It acts by displacing $D$ while keeping the boundary $\Sigma$ fixed. The transformation law for $X$ is given by 
\beq
X \rightarrow X + \ka .
\eeq 
It follows that the Poincar\'e dual $\de_D$ changes by an exact form, $\de_D \rightarrow \de_D + d \lambda$, where the two-form $\lambda$ is given by
\beq
\lambda = \left<d x \wedge d x \right. \int_D \left. \ka \, d X \wedge d X \wedge d X \right> \de^{(6)} (x - X) .
\eeq
(The notation $\left< \ldots \ldots \right>$ here means that the two space-time differentials $d x$, the vector $\ka$, and the three pullback differentials $d X$ are contracted with the Levi-Civita tensor.) To arrive at this expression, we have used Stokes' theorem together with the condition that $\ka$ vanishes on the boundary $\Sigma$ of $D$. We see that $h^{\rm tot}$ is invariant under $\kappa$-symmetry, provided that the let the two-form $b$ transform according to 
\beq
b \rightarrow b - \lambda .
\eeq 
Finally, the last term in $S^{\rm bos}$ is invariant since the pullback of $\lambda$ to $\Sigma$ vanishes. This completes the proof of $\kappa$-symmetry for the bosonic action $S^{\rm bos}$.

\subsection{Off-shell supersymmetry}
We now return to the complete theory including fermions. Off-shell
supersymmetry is largely analogous to off-shell $\kappa$-symmetry: The
statement that $\Phi^{a b}$ and $F$ transform as superfields is only
valid provided that the tensor multiplet fields $\phi^{a b}$,
$\psi^a_\al$, and $h_{\al \be}$ fulfill their free equations of motion
(\ref{EOM}). It follows that the variation of the interaction $S_{NG}
+ S_{WZ}$ when no such constraints are imposed is a linear combination
of $\pa^{\al \be} \pa_{\al \be} \phi$, $\pa^{\al \be} \psi_\be^a$, and
$\pa^{\al \ga} h_{\be \ga}$. It can thus be cancelled by supplementing
the previous supersymmetry transformation laws of the tensor multiplet
fields (\ref{SUSY}) with additional $\de_1$ terms, chosen so that
\beq
\de (S_{NG} + S_{WZ}) + \de_1 S_{TM} = 0 . 
\eeq
But this leads to an additional variation $\de_1 (S_{NG} + S_{WZ})$ of the interaction. We also need to take the variation $\de S_{DD}$ of the additional term $S_{DD}$, which is independent of the tensor multiplet fields, into account. 

The issue of divergences is precisely analogous to the case of
$\ka$-symmetry: The $\de_1$ transformations contain terms supported on
$D$ and on $\Sigma$, and $S_{NG}$ and $S_{WZ}$ are given by integrals
over $\Sigma$ and $D$ respectively. We regulate the ensuing
divergences by instead considering the $\de_1$ variation of
$S_{NG}^\prime$ and $S_{WZ}^\prime$ obtained by using a perturbed
world-volume $D^\prime$ with boundary $\Sigma^\prime$. Also, $S_{DD}$
is regularized to $S_{D^\prime D}$. It follows from the properties of
a generic perturbation $D^\prime$ that $\de_1 S_{NG}^\prime$ vanishes,
so the total variation of the regularized action is
\beq
(\de + \de_1) S = \de_1 S_{WZ}^\prime + \de S_{D^\prime D} ,   
\eeq
where furthermore terms in the $\de_1$ variations of the tensor
multiplet fields that are supported on $\Sigma$ may be neglected in
the computation of $\de_1 S_{WZ}^\prime$. So the right hand side is
given by a double integral over $D$ and $D^\prime$.

A direct proof of the vanishing of the last expression can probably be
found, and is possibly quite illuminating. But in this paper we will
content ourselves with a more implicit argument:  We have shown in the
beginning of this section that the complete action $S$ is invariant
under the $\kappa$-transformations, by which $D$ may be arbitrarily
chosen as long as its boundary is given by the string world-sheet
$\Sigma$. But according to the last equation, the supersymmetry
variation of $S$ is given by a double integral over two copies of $D$
(one of which is actually perturbed to $D^\prime$). We claim that the
integrand of this expression must actually be exact either on $D$ or
on $D^\prime$, so that the integral vanishes by Stokes' theorem. (The
boundary term vanishes, since the intersection of $D$ with
$\Sigma^\prime$ or vice versa is empty.) The reason is that any
non-exact terms would mean that the supersymmetry variation of $S$ had
a dependence on the choice of $D$, which contradicts the fact that $S$
itself is independent of such a choice. This completes the proof of
supersymmetry of the complete action $S$.

To summarize, the complete action is
\beqa
S & = & \int_M d^6 x \left( 2 h_{\al \be} h^{\al \be} - \Om_{a c} \Om_{b d} \pa_{\al \be} \phi^{a b} \pa^{\al \be} \phi^{c d}  - 4 i \Om_{a b} \psi^a_\al \pa^{\al \be} \psi^b_\be \right) \cr
& & {} - \int_\Sigma d^2 \sigma \sqrt{{}^* (\Phi \cdot \Phi)} \sqrt{- G} + \int_D {}^* F - \frac{1}{2} \int_{D^\prime} \int_D {}^* {}^{* \prime} \hat{F}^\prime .
\eeqa
We have shown that it is invariant under a local $\ka$-symmetry, by
which the Dirac membrane world-volume $D$ may be arbitarily chosen as
long as its boundary is given by the string world-sheet $\Sigma$, and
half of the components of $\Theta^\al_a$ on $\Sigma$ may be
eliminated. This transformation acts on the string degrees of freedom
in a standard way, but also on tensor multiplet fields with terms
supported on $D$. Finally, we have shown that the action is invariant
under global $(2, 0)$ supersymmetry transformations, acting on the
string degrees of freedom in the standard way, but with the standard
transformation laws of the tensor multiplet fields supplemented with
terms supported on $D$ and on~$\Si$.

\vspace*{5mm}
M.H. is a Research Fellow at the Royal Swedish Academy of Sciences.


\begin{thebibliography}{99}
\bibitem{Witten95}E.~Witten, {\it Some comments on string dynamics},
  {\tt hep-th/9507121}.
\bibitem{Arvidsson-Flink-Henningson}P.~Arvidsson, E.~Flink and
  M.~Henningson, {\it Thomson scattering of chiral tensors and scalars
    against a self-dual string}, JHEP {\bf 12} (2002) 010, {\tt
    hep-th/0210223}; \\
P.~Arvidsson, E.~Flink and M.~Henningson, {\it Free tensor multiplets
  and strings in spontaneously broken six-dimensional $(2, 0)$
  theory}, JHEP {\bf 06} (2003) 039, {\tt hep-th/0306145}; \\
P.~Arvidsson, E.~Flink and M.~Henningson, {\it Supersymmetric coupling
  of a self-dual string to a $(2, 0)$ tensor multiplet background},
  JHEP {\bf 11} (2003) 015, {\tt hep-th/0309244}.
\bibitem{Witten97}E.~Witten, {\it Five-brane effective action in $M$-theory}, J. Geom. Phys. {\bf 22} (1997) 103, {\tt hep-th/9610234}.
\bibitem{Howe-Sierra-Townsend}P.S.~Howe, G.~Sierra and P.K.~Townsend, {\it Supersymmetry in six dimensions}, Nucl. Phys. {\bf B221} (1983) 331.
\bibitem{Deser-Gomberoff-Henneaux-Teitelboim}S.~Deser, A.~Gomberoff, M.~Henneaux and C.~Teitelboim, {\it p-brane dyons and electric-magnetic duality}, Nucl. Phys. {\bf B520} (1998) 179, {\tt hep-th/9712189}.
\bibitem{Bergshoeff-Sezgin-Townsend}E.~Bergshoeff, E.~Sezgin and P.K.~Townsend, {\it Supermembranes and eleven-dimensional supergravity}, Phys. Lett. {\bf B189} (1987) 75.
\bibitem{Grojean-Mourad}C.~Grojean and J.~Mourad, {\it Superconformal $6 D$ $(2, 0)$ theories in superspace}, Class. Quant. Grav. {\bf 15} (1998) 3397, {\tt hep-th/9807055}.
\bibitem{Howe}P.S.~Howe, {\it Aspects of the $D = 6$, $(2, 0)$ tensor multiplet}, Phys. Lett. {\bf B503} (2001) 197, {\tt hep-th/0008048}.

\end{thebibliography}
\end{document}